\documentclass[pra,twocolumn]{revtex4}
\usepackage{graphicx,amsmath,amssymb,amsthm,hyperref}

\newcommand{\ra}{\rangle}
\newcommand{\la}{\langle}
\def\>{\rangle}
\def\<{\langle}
\def\tr{{\rm Tr}}

\newtheorem{theorem}{Theorem}

\newtheorem{defi}{Definition}

\newcommand{\binomial}[2]{{#1 \choose #2}}

\newcommand{\rar}{\rotatebox[origin=c]{0}{$\circlearrowleft$}}
\newcommand{\lar}{\rotatebox[origin=c]{0}{$\circlearrowright$}}
\newcommand{\eprint}[2][]{\href{http://arxiv.org/abs/#2}{#2}}

\def\bp{\begin{picture}} 
\def\ep{\end{picture}} 
\def\ci{\circle}
\def\li{\line} 

\begin{document}
\title{On independent permutation separability criteria}

\author{Lieven Clarisse}
\email{lc181@york.ac.uk}
\affiliation{Department of Mathematics, The University of York, Heslington, York YO10 5DD, U.K.}
\author{Pawe\l{} Wocjan}
\email{wocjan@cs.caltech.edu}
\affiliation{Computer Science Department \& Institute for
Quantum Information, California Institute of Technology, Pasadena, CA
91125, USA}

\begin{abstract}
Recently P.\ Wocjan and M.\ Horodecki
[\href{http://arxiv.org/abs/quant-ph/0503129}{quant-ph/0503129}] gave
a characterization of combinatorially independent permutation
separability criteria. Combinatorial independence is a necessary
condition for permutations to yield truly independent criteria meaning
that that no criterion is strictly stronger that any other. In this
paper we observe that some of these criteria are still dependent and
analyze why these dependencies occur. To remove them we introduce an
improved necessary condition and give a complete classification of the remaining
permutations. We conjecture that the remaining class of criteria only
contains truly independent permutation separability criteria. Our
conjecture is based on the proof that for two, three and four parties
all these criteria are truly independent and on numerical verification
of their independence for up to $8$ parties. It was
commonly believed that for three parties there were $9$ independent
criteria, here we prove that there are exactly $6$ independent
criteria for three parties and $22$ for four parties.
\end{abstract}

\maketitle

\section{Introduction}
The field of quantum computation and information relies heavily on the
existence of the entanglement phenomenon.  Yet the basic question,
whether a given multipartite quantum state is entangled or not,
remains essentially open. Mathematically, a state is not entangled or
separable if it can be expressed as a convex combination of product
states \cite{Werner89}. For an $r$ party state $\rho$ this means
$$ 
\rho=\sum_i p_i\, \rho^1_i \otimes \rho^2_i \otimes \cdots \otimes
\rho^r_i,
$$
with $p_i>0$ and $\sum_i p_i=1$.  Much work has been done
computationally in devising an efficient algorithm, which would tell
whether the state is entangled or separable. Notable is the work by
Doherty et.\ al. \cite{DPS01,DPS03,DPS04}, where a nested set of
entanglement criteria is constructed, which in the limit of infinite
tests detects every entangled state. When the state is separable, the
algorithm never ends. Thus the convex set of separable states is
iteratively approximated from the outside. The dual approach has been
formulated in Ref.\ \cite{HB04}, which works from the inside,
detecting separability. An independent two-way algorithm for detecting
separability was devised in Ref.\ \cite{EHGC04}. Yet, despite these advances, 
the separability problem is known to be NP-hard \cite{Gurvits03}. For 
the state of the art in these computational approaches the reader is 
referred to Ref.\ \cite{BHH05} and references therein.

Although these computational criteria solve the problem in principle,
they are analytically hard to work with and lack a clear physical
interpretation. Ideally we would also like to have a simple set of
operational criteria detecting most of the entangled states, with a
nice physical interpretation. The prime example of such a criterion is
the partial transpose, originally introduced by Peres
\cite{Peres96}. States not violating this criterion have been shown to
be undistillable or bound entangled \cite{HHH98}. The partial transpose
criterion is basically a transpose operation in one or more of the
parties of the total system. Writing down the quantum state in a
specific basis, this amounts to rearranging the matrix entries. The
cross norm or realignment criterion \cite{Rudolph02, CW02} works
similarly and is independent of the partial transpose criterion. In
particular it can detect bound entangled states. Some analytical
properties of the realignment criterion have been studied in
Ref. \cite{Rudolph02c}. In Ref.\ \cite{HHH02} a unified approach to
these two criteria was presented, which can be extended to
multipartite systems. We refer to these criteria as permutation
criteria.

Let us briefly recall the essential part of these criteria. For
simplicity, consider a bipartite system and the operation $T$ which
performs a transpose of the second subsystem, that is
$$
|i\ra\la j|\otimes |k\ra\la l| \stackrel{T}{\rightarrow} |i\ra\la j|\otimes |l^*\ra\la k^*|.
$$ Where we denote vectors as $|l^*\ra$ for the complex conjugate of
$|l\ra$. Due to linearity, this operation is well defined for
arbitrary quantum states. A state $\rho$ is entangled if the trace
norm $\|T(\rho)\|>1$. Note that the usual partial transpose criterion
says that a state is entangled if $T(\rho)$ has some negative
eigenvalues, but as $T(\rho)$ is Hermitian this is equivalent to
saying that $\|T(\rho)\|>1$. This was first observed in Ref.\
\cite{VW01b}, where it was shown that the quantity $\|T(\rho)\|$ gives
rise to a good entanglement measure.  The realignment criterion
corresponds to the operation $R$ which acts as
$$
|i\ra\la j|\otimes |k\ra\la l| \stackrel{R}{\rightarrow} |i\ra\la k^*|\otimes |j^*\ra\la l|.
$$ Again, we have that a state $\rho$ is entangled if
$\|R(\rho)\|>1$. Other bipartite permutation criteria can be
constructed, but they turn out \cite{HHH02} to be equivalent to either
of these two.

For more than two parties the classification of inequivalent
permutation criteria is an open problem. First steps towards such a
classification have been made in Ref.\ \cite{Fan02, HHH02, WH05b}. For
three parties it was implied that there are $9$
inequivalent criteria \cite{Fan02, WH05b}, and for four parties at
most $34$ inequivalent criteria \cite{WH05b}.

The aim of this paper is to improve upon these results. The paper is
structured as follows. In section~2 we review the work of Wocjan and
Horodecki \cite{WH05b} and introduce their graphical
notation. Section~3 is devoted to our main results; we show that the
class of the so-called combinatorially independent permutation
criteria contains some equivalent criteria which always occur in
pairs.  We completely classify these criteria and give a new upper
bound of the number of independent criteria (this is Theorem~3).  In
particular we find that there are at most $6$ independent criteria for
three parties and $22$ for four parties.  In section~4 we show that
for $2$, $3$ and $4$ parties the criteria from Theorem~3 are truly
independent, in the sense that no criterion is strictly stronger than
any other criterion. Finally, we discuss our results and argue that
there are most likely no more equivalences in the criteria from
Theorem~3.

\section{Notation and previous results}
We start this section with a formal definition of the permutation criteria. Consider an $r$-party state belonging to a Hilbert space $\cal H$, whose subsystems have the same dimension $d$. A general state $\rho\in \cal H$ can be expanded as
$$
\rho=\sum_{i_1,i_2, \ldots, i_{2r}} 
\rho_{i_1 i_2, i_3 i_4, \ldots, i_{2r-1} i_{2r}}
|i_1 i_3\ldots i_{2r-1}\> \<i_2 i_4 \ldots i_{2r}|,
$$
where all indices run from $1$ to $d$.

Let $S_{2r}$ denote the symmetric group with $(2r)!$ elements, that is
the group of the permutations of the set $\{1,2,\ldots, 2r\}$. We
define for each permutation $\sigma\in S_{2r}$ a corresponding map
$\Lambda_{\sigma}$ on operators acting on ${\cal H}$ as
$$
\big[\Lambda_\sigma(\rho)\big]_{i_1 i_2, \ldots, i_{2r-1} i_{2r}}  = 
\rho_{i_{\sigma(1)} i_{\sigma(2)}, \ldots, i_{\sigma(2r-1)} i_{\sigma(2r)}}\,.
$$ We will represent permutations as $[\sigma(1)\, \sigma(2)\,
\ldots\, \sigma (2r)]$ or in disjoint cycles (an example below will
make this clear).  In Ref.\ \cite{HHH02} it was shown that every
permutation $\sigma\in S_{2r}$ gives rise to an entanglement
criterion. Namely, a state $\rho$ is entangled, if
$\|\Lambda_\sigma(\rho)\|>1$ for any permutation $\sigma\in S_{2r}$,
where $\|A\|=\tr (AA^\dagger)^{1/2}$ denotes the trace norm.

Let us illustrate these definitions for bipartite quantum states. With
the notation introduced above the partial transpose criterion
corresponds to the permutation $[1\,2\,4\,3]=(3,\,4)$ while the
realignment criterion corresponds to the permutation
$[1\,3\,2\,4]=(2,\,3)$.

\begin{defi}[Independent permutation criteria]
Let $\sigma$ and $\mu$ be two permutations in $S_{2r}$. We call the
corresponding entanglement criteria $\sigma$ and $\mu$  dependent
if and only if
\begin{equation}\label{eq:equalNorm}
\| \Lambda_\sigma(\rho)\| = \|\Lambda_\mu(\rho)\|,
\end{equation}
for all quantum states (that is, positive operators with trace
$1$). Else, the permutation criteria are called  independent.
\end{defi}

This definition is motivated by the fact that independence is a
necessary condition for two permutations to yield truly independent
entanglement criteria. This condition extends the necessary condition
of combinatorial independence introduced in \cite{WH05b}. The
definition of combinatorial independence is very similar to the
definition of independence. The decisive difference is that two
permutations are combinatorially dependent if and only if equality in
(\ref{eq:equalNorm}) is achieved for {\em all} operators and not only
quantum states. In this case, the maps $\Lambda_\sigma$ and
$\Lambda_\mu$ are related by a norm-preserving map. A norm preserving
map $\Lambda$ is a map such that for any operator $A$ (not only
density operators), $\| A\|=\|\Lambda(A)\|$. Moreover, it can be shown
that if $\Lambda_\sigma$ and $\Lambda_\mu$ are related via a
norm-preserving map $\Lambda$, then $\Lambda=\Lambda_\nu$ must come
from some permutation $\nu$. We call such a permutation
norm-preserving permutation.
 
An example of such a norm-preserving permutation is the global quantum
transpose (GQT), which transposes the complete system. It can be
written as $\tau=(1,2)(3,4)\cdots (2r-1,2r)$. Another example of a
norm preserving map $\Lambda$ is the ``unitary'' transformation
$\Lambda(\rho)=U\rho V$, where $U$ and $V$ are unitary
operators. Reordering the different parties in the density matrix
representation is an example of such a ``unitary''
transformation. Consider for instance the transformation
$$
\rho=\sum_i \rho^A_i \otimes \rho^B_i \rightarrow \rho'=\sum_i \rho^B_i \otimes \rho^A_i.
$$ This mapping can be implemented by means of multiplication on the
left and on the right by the swap operator \cite{WH05b}.

To illustrate the necessary condition of combinatorial independence
let us consider again a bipartite system. The operation $R'$ induced
by the permutation $(1,\,4)$ gives rise to the same criterion as $R$
induced by the permutation $(2,\,3)$. Indeed, the permutation $R'\tau$
(here and elsewhere permutations are evaluated from \emph{left} to
\emph{right}) transforms $[1\,2\,3\,4]$ into $[2\,4\,1\,3]$, which up
to reordering of the parties is equivalent to $[1\,3\,2\,4]$. This is
just the transformation defined by $R$.

\begin{theorem}[Combinatorially independent criteria \cite{WH05b}]
The group $\cal T$ of norm preserving permutations can be generated as
$$
{\cal T}= \la(2k,2l),(2k-1,2l-1),\tau \ra,
$$ 
where $1\leq k,l\leq r$ and $\tau$ as before denotes the GQT. The
combinatorially independent permutation criteria correspond to the
right cosets $S_{2r}/{\cal T}$ of ${\cal T}$ in $S_{2r}$. The number
of independent permutation criteria is therefore not larger than
$ {\frac{1}{2}}\binomial{2r}{r}.$ 
In this number, the class of trivial norm preserving criteria is
also counted.
\end{theorem}
In the same paper, the authors also devised a graphical notion of the
criteria, which leads to a way of selecting a simple representative
for the right cosets. They decompose any permutation as a combination
of 4 elementary permutations: the partial transpose, two types of realignment 
or reshuffling, and the identity. The corresponding graphical
notations are loops from a subsystem to itself, arrows from one
subsystem to another and free subsystems  (no loops or arrows), as graphically depicted in
Figure~\ref{tab:basicPermutations}. We call $k$ the head and $l$ the
tail if there is an arrow from $k$ to $l$. If there is a loop in $m$,
then $m$ is both head and tail of the loop. We call the support of an
arrow or a loop the set containing its head and tail. A configuration
of arrows and loops is called disjoint, if the supports of all pairs
and loops are disjoint.
\begin{figure*}
\begin{tabular}{c|c|c}
\quad\quad graphical representation\quad\quad &
\quad\quad corresponding permutation\quad\quad & 
\quad\quad name \quad\quad \\ \hline\hline
$k\quad \bullet\longrightarrow\bullet\quad l$ & $(2k,\, 2l-1)$ & reshuffle 
$R_{kl}$ \\\hline
$k\quad\bullet\longleftarrow \bullet\quad l$ & $(2k-1,\, 2l)$ & reshuffle 
$R'_{lk}$ \\\hline
$k\quad\lar$\quad\quad\quad\quad\quad        & $(2k-1,\, 2k)$ &  \quad 
partial transpose\quad\\\hline
$k\quad\bullet$\quad\quad\quad\quad\quad     & ()          & identity
\end{tabular}
\caption{Basic permutations (Table from Ref. \cite{WH05b}).}
\label{tab:basicPermutations}
\end{figure*}

\begin{theorem}[Canonical representation of combinatorially independent criteria \cite{WH05b}]
The right cosets $S_{2r}/{\cal T}$ can always be represented by a
disjoint configuration of arrows. All criteria that can be represented
in this manner are combinatorially independent up to reversing the
direction of all arrows and replacing loops by free subsystems and
vice versa.
\end{theorem}

\section{Main Results}
It was shown in Ref.\ \cite{WH05b} that two permutations $\mu$ and
$\sigma$ are combinatorially dependent if there is a norm preserving
permutation such $\nu$ that $\Lambda_\sigma(A) =
\Lambda_\nu(\Lambda_\mu(A))$ for all operators $A$. One could assume
that therefore Theorem~1 cannot be sharpened, that is, all
combinatorially independent criteria are independent. This is true
when one considers the permutation acting on arbitrary operators. In
quantum mechanics however, we deal with positive operators, which are
Hermitian. The following theorem exploits this fact and counts the new
upper bound to the number of independent criteria.

\begin{figure*}
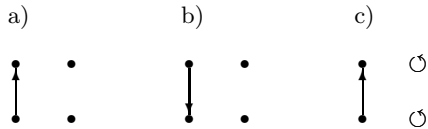

\vspace{1cm}
\unitlength=2pt
\begin{tabular}{lll}
a) & b) & c) \\ && \\ && \\
\,
\begin{minipage}{2cm}
\bp(0,0)(15,0) 
\put(0,1){\vector(0,1){8.5}}
\put(0,10.5){\ci*{1.5}}
\put(10.5,10.5){\ci*{1.5}}
\put(0,0){\ci*{1.5}}
\put(10.5,0){\ci*{1.5}}\ep 
\end{minipage}
&
\,
\begin{minipage}{2cm}
\bp(0,0)(15,0) 
\put(0,9.5){\vector(0,-1){8.5}}
\put(0,10.5){\ci*{1.5}}
\put(10.5,10.5){\ci*{1.5}}
\put(0,0){\ci*{1.5}}
\put(10.5,0){\ci*{1.5}}\ep 
\end{minipage}
&
\,
\begin{minipage}{2cm}
\bp(0,0)(15,0) 
\put(0,1){\vector(0,1){8.5}}
\put(0,10.5){\ci*{1.5}}
\put(10.5,10.5){\makebox(0,0){$\rar$}} 
\put(10.55,10.5){\makebox(0,0){$\rar$}} 
\put(10.5,10.55){\makebox(0,0){$\rar$}} 
\put(0,0){\ci*{1.5}}
\put(10.5,0){\makebox(0,0){$\rar$}} 
\put(10.55,0){\makebox(0,0){$\rar$}} 
\put(10.5,0.05){\makebox(0,0){$\rar$}} 
\ep 
\end{minipage}
\end{tabular}
\caption{Action of the new rule on a disjoint arrow configuration: the
first equivalence corresponds to the new rule and the second corresponds to
Rule~4 in \cite{WH05b}.}
\label{fig:actionNewRule}
\end{figure*}

\begin{theorem}
(i). The permutation criteria corresponding to the permutations
$\sigma$ and $\tau\sigma$ are dependent. (ii). Let ${\cal
Z}:=\{e,\tau\}$ be the subgroup of $S_{2r}$ generated by the QGT
$\tau$. Define the action of $\cal Z$ on the right cosets
$S_{2r}/{\cal T}$ by multiplication from the left of the cosets, that
is, $e*\sigma {\cal T} = e\sigma \cal T$ and $\tau*\sigma {\cal T} =
\tau\sigma \cal T$. The new upper bound on the number of independent
criteria is the number of orbits under this action. It is given by
\footnote{The integer sequence generated by the number of independent
criteria equals the integer sequence A072377 from Ref.~\cite{sloan}.}
\begin{equation}
\label{uupp}
\frac{1}{4}\left[\binomial{2r}{r} + 2^r + \binomial{r}{r/2}\,
\cdot\, even(r)\right]\,,
\end{equation}
where $even(r)=1$ if $r$ is even and $0$ otherwise. This number
includes the trivial criterion given by the identity permutation.
\end{theorem}
\begin{proof}
To prove (i), let us apply the permutation $\tau\sigma$ on an
arbitrary quantum state $\rho$. We have
\begin{align*}
\|\Lambda_{\tau \sigma}(\rho)\|&=\|\Lambda_{\sigma}(\Lambda_{\tau }(\rho))\|\\
&=\|\Lambda_{\sigma}(\rho^T)\| \\
&=\|\Lambda_{\sigma}(\bar{\rho})\| \\
&=\|\overline{\Lambda_{\sigma}(\rho)}\|\\
&=\|\Lambda_{\sigma}(\rho)\|.
\end{align*}
Here we have used that $\rho^T=\bar{\rho}$ because $\rho$ is Hermitian
and $\Lambda_{\sigma}(\bar{\rho})=\overline{\Lambda_{\sigma}(\rho)}$
because $\Lambda$ only permutes the entries of the matrix.

Observe that multiplying a coset $\sigma \cal T$ by $\tau$ from the
left is the same as conjugating it by $\tau$ because $\tau$ is
contained in $\cal T$, that is, we have
$$
\tau\sigma \cal T = \tau \sigma \cal T \tau.
$$ It is readily verified that conjugation of a permutation by $\tau$
corresponds to exchanging heads (always odd numbers) and tails (always
even numbers). Therefore, the direction of all (true) arrows is
reversed and loops and free subsystems are not affected. An example is
shown in Fig.~\ref{fig:actionNewRule}. Following the four rules
presented in \cite{WH05b}, we call this Rule 5.  The new rule either
\emph{glues} two criteria together or does not change them. More
precisely, the orbits under the action of $\cal Z$ have size $1$ or
$2$. This is because $\tau$ is an involution
$$
\sigma\cal T \rightarrow \tau\sigma\cal T \rightarrow \tau\tau\sigma\cal T=\sigma\cal T.
$$

(ii). There are $\frac{1}{2}\binomial{2r}{r}$ combinatorial
independent criteria. With the new rule, there are at most
 $$
\frac{1}{2}\binomial{2r}{r} -
\frac{1}{2}\left[\frac{1}{2}\binomial{2r}{r}-K\right],
$$
independent criteria left. Here $K$ denotes the number of criteria not affected by conjugating by
the new rule. Now note that the only criteria (represented as disjoint
arrow configurations) not affected by conjugation with $\tau$ are
\begin{enumerate}
\item criteria with no arrows and
\item criteria containing only arrows and having no free subsystems.
\end{enumerate}
If $r$ is odd, then situation (2) cannot occur. The number of these
criteria are readily counted:
$$
\binomial{r}{0}+ \binomial{r}{1}+\binomial{r}{2}+\cdots+ 
\binomial{r}{\lfloor(r/2)\rfloor}=2^{r-1},
$$ where we have used an identity of binomial coefficients.  So that
in the case of an odd number of subsystems, the number of 
criteria becomes (including the identity)
$$
\frac{1}{4}\left[\binomial{2r}{r}+2^r\right].
$$

In the case $r$ is even we need to take care of situation (2). Now this number equals picking $r/2$ heads from $r$ choices, because exchanging tails does not matter. But we have to divide by two since exchanging all heads and tails does not matter either, so that the number of criteria satisfying (2) is given by
$$
\frac{1}{2}\left[\binomial{r}{r/2}\right].
$$ We conclude that in the case $r$ is even, the number of
criteria is given by (including the identity)
$$
\frac{1}{4}\left[\binomial{2r}{r} + 2^r + \binomial{r}{r/2} \right].
$$ To complete the proof, we have to show that the criteria in the
orbits of size $2$ are combinatorially independent. Let $\sigma$ be a
permutation represented by a disjoint arrow configuration. Assume that
there is an arrow from subsystem $k$ to $l$ in the disjoint
configuration of $\sigma$. Then there is an arrow from $l$ to $k$ in
the disjoint configuration describing $\tau\sigma\tau$. Loops and free
subsystems are not affected. Using these observation we see that the
configuration describing $\sigma\tau\sigma\tau$ has a closed path from
$k$ to $l$ and no loops. The closed path between $k$ and $l$ can be
transformed into a loop on $k$ and a loop on $l$ with the help of
Rule~3 (Exchanging heads) in Ref.~\cite{WH05b}. Now if we apply these
arguments to all arrows of $\sigma$ we see that the permutation
$\sigma\tau\sigma\tau$ is not norm-preserving. Consequently, the
permutations $\sigma$ and $\tau\sigma\tau$ are combinatorial
independent. This concludes the proof.
\end{proof}

Fig.~\ref{fig:actionNewRule} shows an example of how the new rule
glues two combinatorially independent criteria together. In general, we
have that two permutations corresponding to disjoint arrow
configurations are related by the new rule if and only if they have the same
(true) arrow structure (up to exchanging heads) and a complementary
loop/free subsystem structure. The latter means that if the first
criterion has a loop on subsystem $k$ then the subsystem $k$ is free
in the second and vice versa.

\section{Illustrations}
In this section we will illustrate the permutation criteria for two, three and four parties. The different criteria are shown graphically in Figure~\ref{fig:four}. Here loops (partial
transpose) are depicted by a little circle, solid lines represent the
first type of reshuffling and dotted lines, the second type of
reshuffling. In this section we go further and show that the 
criteria from Theorem~3 are \emph{truly} independent: no criteria detects strictly
more states than any other criteria.

\begin{figure*}
\begin{tabular}{rl}
a) & \\ &
\begin{tabular}{c|l}
\unitlength=1pt
&\\[-2mm]
\begin{minipage}{1.2cm} QT \end{minipage} & \
\ $\bp(0,0)(0,-5) \put(0,0){\ci{3}}\put(10.5,0){\ci*{2}}\ep $
\quad\quad\ \  
\\[2mm]
\hline
&\\[-2mm]
R & \
\ $\bp(0,0)(0,-5) \put(0,0){\li(1,0){10.1}}
\put(0,0){\ci*{2}}\put(10.5,0){\ci*{2}}\ep $
\quad\quad\  \  
\end{tabular}
\\\\
b) & \\ &
\begin{tabular}{c|l}
\unitlength=1pt
&\\[-2mm]
\begin{minipage}{1.2cm} QT \end{minipage} & \
\ $\bp(0,0)(0,1) \put(0,10.5){\ci{3}}\put(0,0){\ci*{2}}\put(10.5,0){\ci*{2}}\ep $
\quad\quad\ \  
\ $\bp(0,0)(0,1) \put(0,10.5){\ci*{2}}\put(0,0){\ci{3}}\put(10.5,0){\ci*{2}}\ep $
\quad\quad\ \  
\ $\bp(0,0)(0,1) \put(0,10.5){\ci*{2}}\put(0,0){\ci*{2}}\put(10.5,0){\ci{3}}\ep $
\quad\quad\ \  
\\[2mm]
\hline
&\\[-2mm]
R & \
\ $\bp(0,0)(0,1) \put(0,0){\li(0,1){10.1}}
\put(0,10.5){\ci*{2}}\put(0,0){\ci*{2}}\put(10.5,0){\ci*{2}}\ep $
\quad\quad\  \  
\ $\bp(0,0)(0,1) \put(0,0){\li(1,0){10.1}}
\put(0,10.5){\ci*{2}}\put(0,0){\ci*{2}}\put(10.5,0){\ci*{2}}\ep $
\quad\quad\  \  
\ $\bp(0,0)(0,1) \put(0,10.5){\li(1,-1){10}}
\put(0,10.5){\ci*{2}}\put(0,0){\ci*{2}}\put(10,0){\ci*{2}}\ep $
\quad\quad\  \  
\end{tabular}
\\\\
c) & \\ &
\begin{tabular}{c|l}
\unitlength=1pt
&\\[-2mm]
\begin{minipage}{1.2cm} QT \end{minipage} & \
\ $\bp(0,0)(0,1) \put(0,10.5){\ci{3}}\put(10.5,10.5){\ci*{2}}\put(0,0){\ci*{2}}\put(10.5,0){\ci*{2}}\ep $
\quad\quad\ \  
\ $\bp(0,0)(0,1) \put(0,10.5){\ci*{2}}\put(10.5,10.5){\ci{3}}\put(0,0){\ci*{2}}\put(10.5,0){\ci*{2}}\ep $
\quad\quad\ \  
\ $\bp(0,0)(0,1) \put(0,10.5){\ci*{2}}\put(10.5,10.5){\ci*{2}}\put(0,0){\ci{3}}\put(10.5,0){\ci*{2}}\ep $
\quad\quad\ \  
\ $\bp(0,0)(0,1) \put(0,10.5){\ci*{2}}\put(10.5,10.5){\ci*{2}}\put(0,0){\ci*{2}}\put(10.5,0){\ci{3}}\ep $
\quad\quad\ \  
\\[2mm]
\hline
&\\[-2mm]
2QT & \
\ $\bp(0,0)(0,1) \put(0,10.5){\ci{3}}\put(10.5,10.5){\ci*{2}}\put(0,0){\ci{3}}\put(10.5,0){\ci*{2}}\ep $
\quad\quad\  \  
\ $\bp(0,0)(0,1) \put(0,10.5){\ci{3}}\put(10.5,10.5){\ci{3}}\put(0,0){\ci*{2}}\put(10.5,0){\ci*{2}}\ep $
\quad\quad\ \  
\ $\bp(0,0)(0,1) \put(0,10.5){\ci{3}}\put(10.5,10.5){\ci*{2}}\put(0,0){\ci*{2}}\put(10.5,0){\ci{3}}\ep $
\quad\quad\ \  
\\[2mm]
\hline
&\\[-2mm]
R & \
\ $\bp(0,0)(0,1) \put(0,0){\li(0,1){10.1}}
\put(0,10.5){\ci*{2}}\put(10.5,10.5){\ci*{2}}\put(0,0){\ci*{2}}\put(10.5,0){\ci*{2}}\ep $
\quad\quad\  \  
\ $\bp(0,0)(0,1) \put(10.5,0){\li(0,1){10.1}}
\put(0,10.5){\ci*{2}}\put(10.5,10.5){\ci*{2}}\put(0,0){\ci*{2}}\put(10.5,0){\ci*{2}}\ep $
\quad\quad\  \  
\ $\bp(0,0)(0,1) \put(0,0){\li(1,0){10.1}}
\put(0,10.5){\ci*{2}}\put(10.5,10.5){\ci*{2}}\put(0,0){\ci*{2}}\put(10.5,0){\ci*{2}}\ep $
\quad\quad\  \  
\ $\bp(0,0)(0,1) \put(0,10.5){\li(1,0){10.1}}
\put(0,10.5){\ci*{2}}\put(10.5,10.5){\ci*{2}}\put(0,0){\ci*{2}}\put(10.5,0){\ci*{2}}\ep $
\quad\quad\  \  
\ $\bp(0,0)(0,1) \put(0,0){\li(1,1){10}}
\put(0,10.5){\ci*{2}}\put(10.5,10.5){\ci*{2}}\put(0,0){\ci*{2}}\put(10,0){\ci*{2}}\ep $
\quad\quad\  \  
\ $\bp(0,0)(0,1) \put(0,10.5){\li(1,-1){10}}
\put(0,10.5){\ci*{2}}\put(10.5,10.5){\ci*{2}}\put(0,0){\ci*{2}}\put(10,0){\ci*{2}}\ep $
\quad\quad\  \  
\\[2mm]
\hline
&\\[-2mm]
R+QT &  \
\ $\bp(0,0)(0,1) \put(0,0){\li(0,1){10.1}}
\put(0,10.5){\ci*{2}}\put(10.5,10.5){\ci*{2}}\put(0,0){\ci*{2}}\put(10.5,0){\ci{3}}\ep $
\quad\quad\  \  
\ $\bp(0,0)(0,1) \put(10.5,0){\li(0,1){10.1}}
\put(0,10.5){\ci{3}}\put(10.5,10.5){\ci*{2}}\put(0,0){\ci*{2}}\put(10.5,0){\ci*{2}}\ep $
\quad\quad\  \  
\ $\bp(0,0)(0,1) \put(0,0){\li(1,0){10.1}}
\put(0,10.5){\ci*{2}}\put(10.5,10.5){\ci{3}}\put(0,0){\ci*{2}}\put(10.5,0){\ci*{2}}\ep $
\quad\quad\  \  
\ $\bp(0,0)(0,1) \put(0,10.5){\li(1,0){10.1}}
\put(0,10.5){\ci*{2}}\put(10.5,10.5){\ci*{2}}\put(0,0){\ci{3}}\put(10.5,0){\ci*{2}}\ep $
\quad\quad\  \  
\ $\bp(0,0)(0,1) \put(0,0){\li(1,1){10}}
\put(0,10.5){\ci{3}}\put(10.5,10.5){\ci*{2}}\put(0,0){\ci*{2}}\put(10,0){\ci*{2}}\ep $
\quad\quad\  \  
\ $\bp(0,0)(0,1) \put(0,10.5){\li(1,-1){10}}
\put(0,10.5){\ci*{2}}\put(10.5,10.5){\ci*{2}}\put(0,0){\ci{3}}\put(10,0){\ci*{2}}\ep $
\quad\quad\  \  
\\[2mm]
\hline
&\\[-2mm]
2R & \
\ \bp(0,0)(0,1) \put(0,0){\li(0,1){10.1}}\put(10.5,0){\li(0,1){10.1}}
\put(0,10.5){\ci*{2}}\put(10.5,10.5){\ci*{2}}\put(0,0){\ci*{2}}\put(10.5,0){\ci*{2}}\ep 
\quad\quad\  \  
\ \bp(0,0)(0,1) \put(0,0){\li(1,0){10.1}}\put(0,10.5){\li(1,0){10.1}}
\put(0,10.5){\ci*{2}}\put(10.5,10.5){\ci*{2}}\put(0,0){\ci*{2}}\put(10.5,0){\ci*{2}}\ep 
\quad\quad\  \  
\\[2mm]
\hline
&\\[-2mm]
R+R' & \ 
\ $\bp(0,0)(0,1) \put(0,0){\li(0,1){10.1}}
\multiput(10.5,0.25)(0,2){6}{\ci*{0.5}}
\put(0,10.5){\ci*{2}}\put(10.5,10.5){\ci*{2}}\put(0,0){\ci*{2}}\put(10.5,0){\ci*{2}}\ep $
\quad\quad\  \  \\
\end{tabular}
\end{tabular}
\caption[four]{Independent permutation criteria for
a) two, b) three, and c) four particles. Picture adapted from Ref.\ \cite{WH05b}.}
\label{fig:four}
\end{figure*}

\subsection{Two parties}
For a quantum system consisting of two parties, there are only two non
trivial inequivalent permutation criteria: the partial transpose in
one of the subsystems and reshuffling between the two subsystems. For
the low dimensional systems ${\cal H}\cong \mathbb{C}^2 \otimes
\mathbb{C}^2$ and ${\cal H}\cong \mathbb{C}^2 \otimes \mathbb{C}^3$
the positivity of the partial transpose is a necessary and sufficient
condition for separability \cite{HHH96}, while this is not the case
for the realignment criterion \cite{Rudolph02c, CW02}. For higher
dimensional systems these criteria are truly independent. We have
tested the realignment criterion on all known bound entangled states
$\rho \in{\cal H}\cong \mathbb{C}^3 \otimes \mathbb{C}^3$ in the
literature. The maximum value we have found for $\|R(\rho)\|$ is $7/6$
and is achieved for a particular chess-board state \cite{BP99}:
$$
\rho_c=\frac{1}{12}\left[ \begin{array}{rrrrrrrrr}
  1 & 0 & 1 & 0 & 0 & 0 & \,\,\,\, 1 & 0 & 0 \\
  0 & 1 & 0 & 0 & 0 & -1 & 0 & -1 & 0 \\
  1 & 0 & 2 & 0 & -1 & 0 & 0 & 0 & 0 \\
  0 & 0 & 0 & 1 & 0 & -1 & 0 & 1 & 0 \\
  0 & 0 & -1 & 0 & 1 & 0 & 1 & 0 & 0 \\
  0 & -1 & 0 & -1 & 0 & 2 & 0 & 0 & 0 \\
  1 & 0 & 0 & 0 & 1 & 0 & 2 & 0 & 0 \\
  0 & -1 & 0 & 1 & 0 & 0 & 0 & 2 & 0 \\
  0 & 0 & 0 & 0 & 0 & 0 & 0 & 0 & 0
\end{array} \right].
$$

\subsection{Three parties}
For three parties we have proven that only $6$ criteria are
independent, previous work \cite{WH05b, Fan02} indicated that there
were $9$. The $6$ criteria are the partial transposes (row QT) in the
$3$ subsystems and the $3$ reshufflings (row R) between any of the two
subsystems.

To show that all the criteria from row QT are independent, it is
sufficient to note that there exist tripartite states which are
separable with regard to two splits, but not to the third one (for an
example with qubits, see Ref. \cite{DC99}). It has been proven
\cite{Rudolph02c} that the trace norm of the realigned density matrix
remains invariant when an uncorrelated ancilla is added. Now take a
bipartite entangled state which violates the realignment criterion but
not the partial transpose criterion (such as $\rho_c$). By adding an
uncorrelated ancilla and reordering the systems, we can construct
states that are detected by one criterion from row R only. These
states are trivially not detected by any criterion from row QT as they
are bound entangled.

Note that the realignment criterion, in contrast to the partial
transpose criterion can detect genuine tripartite entangled states,
that is, entangled states that are separable under any bipartite
cut. This has been demonstrated in Ref.\ \cite{HHH02} with the
tripartite bound entangled states from Ref.\ \cite{BDMSST99}.

\subsection{Four parties}
For four parties, there are at 22 non trivial independent permutation
criteria. As for three parties, it is trivial to construct states that
are only detected by one partial transpose criterion (that is a
criterion with only loops). Again, each criterion with at least one
realignment is truly independent of the transpose criteria because the
trace norm of the realigned density matrix remains invariant when an
uncorrelated ancilla is added.

To show true independence within the set of realignment criteria (rows
R, R+QT, 2R and R+R'), let us first consider the rows R and
R+QT. Using states with a negative partial transpose, it is very easy
to construct examples to show that these criteria are independent from
each other and from the rows R and R'. We verified this using a random
search over the state space on 4 qubits (using the algorithm outlined
in Ref.\ \cite{ZHSL98}). In the same way it can be checked that the
criteria from the rows 2R and R+R' are independent from each other. To
show that they are also independent from the rows R and R+QT, it can
be verified that they detect states of the form
$\rho=(1-\beta)\rho_c\otimes \rho_c+\beta \openone/81$ for a larger
parameter range of $\beta>0$.

\section{Discussion}
In Ref.\ \cite{HHH02} a powerful class of separability criteria was
devised based on permutations. The class however contained many
redundancies, and to give a complete characterization of the
independent criteria is an open problem. In Ref.\ \cite{WH05b} a
graphical representation for permutations together with rules for
simplifying them were introduced based on a sufficient condition for
two permutations to yield dependent criteria. This equivalence meant
that two combinatorially dependent criteria yield the same value of
the trace norm on all operators. Combinatorially, density operators
have a prominent Hermitian symmetry, that is a global quantum
transposition together with a complex conjugate.  In this paper we
have exploited this symmetry and we have shown how this lead to
dependence of particular combinatorially independent criteria. It is
unlikely that there are more dependences in the criteria from
Theorem~3.

 Density operators differ from arbitrary operators also in
the fact that they have positive eigenvalues. But since permutations
only reorder matrix entries, it is not very likely that this
positiveness would lead to more criteria to be dependent. We have verified 
the independence of the criteria numerically on a random state for 2 upto 8 parties.

In the same way as we illustrated for three and four parties, it is
easy to see that the criteria with only loops (only partial
transpositions) are independent. These criteria are independent from
the ones having at least one realignment since those can detect bound
entangled states. To prove the independence within the class of
criteria with at least one realignment one could try to generalize the
arguments from Section 4. 

\begin{acknowledgments}
L.~C.\ is supported by a WW Smith Scholarship. P.~W.\ is supported by
the National Science Foundation under the grant no.~EIA
0086038. L.~C.\ would like to thank Anthony Sudbery for careful
reading of the manuscript and helpful suggestions. L.~C.\ also thanks
Christine Aronsen Storeb{\o} for very helpful discussions.
\end{acknowledgments}


\begin{thebibliography}{23}
\expandafter\ifx\csname natexlab\endcsname\relax\def\natexlab#1{#1}\fi
\expandafter\ifx\csname bibnamefont\endcsname\relax
  \def\bibnamefont#1{#1}\fi
\expandafter\ifx\csname bibfnamefont\endcsname\relax
  \def\bibfnamefont#1{#1}\fi
\expandafter\ifx\csname citenamefont\endcsname\relax
  \def\citenamefont#1{#1}\fi
\expandafter\ifx\csname url\endcsname\relax
  \def\url#1{\texttt{#1}}\fi
\expandafter\ifx\csname urlprefix\endcsname\relax\def\urlprefix{URL }\fi
\providecommand{\bibinfo}[2]{#2}
\providecommand{\eprint}[2][]{\url{#2}}

\bibitem[{\citenamefont{Werner}(1989)}]{Werner89}
\bibinfo{author}{\bibfnamefont{R.~F.} \bibnamefont{Werner}},
  \emph{\bibinfo{title}{Quantum states with {E}instein-{P}odolsky-{R}osen
  correlations admitting a hidden-variable model}}, \bibinfo{journal}{Physical
  Review A} \textbf{\bibinfo{volume}{40}}, \bibinfo{pages}{4277}
  (\bibinfo{year}{1989}).

\bibitem[{\citenamefont{Doherty et~al.}(2002)\citenamefont{Doherty, Parrilo,
  and Spedalieri}}]{DPS01}
\bibinfo{author}{\bibfnamefont{A.~C.} \bibnamefont{Doherty}},
  \bibinfo{author}{\bibfnamefont{P.~A.} \bibnamefont{Parrilo}},
  \bibnamefont{and} \bibinfo{author}{\bibfnamefont{F.~M.}
  \bibnamefont{Spedalieri}}, \emph{\bibinfo{title}{Distinguishing separable and
  entangled states}}, \bibinfo{journal}{Physical Review Letters}
  \textbf{\bibinfo{volume}{88}}, \bibinfo{pages}{187904}
  (\bibinfo{year}{2002}), \eprint{quant-ph/0112007}.

\bibitem[{\citenamefont{Doherty et~al.}(2004)\citenamefont{Doherty, Parrilo,
  and Spedalieri}}]{DPS03}
\bibinfo{author}{\bibfnamefont{A.~C.} \bibnamefont{Doherty}},
  \bibinfo{author}{\bibfnamefont{P.~A.} \bibnamefont{Parrilo}},
  \bibnamefont{and} \bibinfo{author}{\bibfnamefont{F.~M.}
  \bibnamefont{Spedalieri}}, \emph{\bibinfo{title}{Complete family of
  seperability criteria}}, \bibinfo{journal}{Physical Review A}
  \textbf{\bibinfo{volume}{69}}, \bibinfo{pages}{022308}
  (\bibinfo{year}{2004}), \eprint{quant-ph/0308032}.

\bibitem[{\citenamefont{Doherty et~al.}(2005)\citenamefont{Doherty, Parrilo,
  and Spedalieri}}]{DPS04}
\bibinfo{author}{\bibfnamefont{A.~C.} \bibnamefont{Doherty}},
  \bibinfo{author}{\bibfnamefont{P.~A.} \bibnamefont{Parrilo}},
  \bibnamefont{and} \bibinfo{author}{\bibfnamefont{F.~M.}
  \bibnamefont{Spedalieri}}, \emph{\bibinfo{title}{Detecting multipartite
  entanglement}}, \bibinfo{journal}{Physical Review A}
  \textbf{\bibinfo{volume}{71}}, \bibinfo{pages}{032333}
  (\bibinfo{year}{2005}), \eprint{quant-ph/0407143}.

\bibitem[{\citenamefont{Hulpke and Bru{\ss}}(2004)}]{HB04}
\bibinfo{author}{\bibfnamefont{F.}~\bibnamefont{Hulpke}} \bibnamefont{and}
  \bibinfo{author}{\bibfnamefont{D.}~\bibnamefont{Bru{\ss}}},
  \emph{\bibinfo{title}{A two-way algorithm for the entanglement problem}}
  (\bibinfo{year}{2004}), \eprint{quant-ph/0407179}.

\bibitem[{\citenamefont{Eisert et~al.}(2004)\citenamefont{Eisert, Hyllus, Guehne,
  and Curty}}]{EHGC04}
\bibinfo{author}{\bibfnamefont{J.} \bibnamefont{Eisert}},
  \bibinfo{author}{\bibfnamefont{P.} \bibnamefont{Hyllus}},
  \bibinfo{author}{\bibfnamefont{O.} \bibnamefont{G{\"u}hne}},
  \bibnamefont{and} \bibinfo{author}{\bibfnamefont{M.}
  \bibnamefont{Curty}}, \emph{\bibinfo{title}{Complete hierarchies of efficient approximations
 to problems in entanglement theory}}, \bibinfo{journal}{Physical Review A}
  \textbf{\bibinfo{volume}{70}}, \bibinfo{pages}{062317}
  (\bibinfo{year}{2004}), \eprint{quant-ph/0407135}.

\bibitem[{\citenamefont{Gurvits}(2003)}]{Gurvits03}
\bibinfo{author}{\bibfnamefont{L.}~\bibnamefont{Gurvits}}, in
  \emph{\bibinfo{booktitle}{Proceedings of the 35 ACM symposium on Theory of
  computing}} (\bibinfo{address}{New York}, \bibinfo{year}{2003}), pp.
  \bibinfo{pages}{10--19 (see
  {\href{http://arxiv.org/abs/quant--ph/0303055}{quant--ph/0303055}} for the
  long version)}.

\bibitem[{\citenamefont{Badzi{\c{a}}g et~al.}(2005)\citenamefont{Badzi{\c{a}}g,
  Horodecki, and Horodecki}}]{BHH05}
\bibinfo{author}{\bibfnamefont{P.}~\bibnamefont{Badzi{\c{a}}g}},
  \bibinfo{author}{\bibfnamefont{P.}~\bibnamefont{Horodecki}},
  \bibnamefont{and}
  \bibinfo{author}{\bibfnamefont{R.}~\bibnamefont{Horodecki}},
  \emph{\bibinfo{title}{Towards efficient algorithm deciding separability of
  distributed quantum states}} (\bibinfo{year}{2005}),
  \eprint{quant-ph/0504041}.

\bibitem[{\citenamefont{Peres}(1996)}]{Peres96}
\bibinfo{author}{\bibfnamefont{A.}~\bibnamefont{Peres}},
  \emph{\bibinfo{title}{Separability criterion for density matrices}},
  \bibinfo{journal}{Physical Review Letters} \textbf{\bibinfo{volume}{77}},
  \bibinfo{pages}{1413} (\bibinfo{year}{1996}), \eprint{quant-ph/9604005}.

\bibitem[{\citenamefont{Horodecki et~al.}(1998)\citenamefont{Horodecki,
  Horodecki, and Horodecki}}]{HHH98}
\bibinfo{author}{\bibfnamefont{M.}~\bibnamefont{Horodecki}},
  \bibinfo{author}{\bibfnamefont{P.}~\bibnamefont{Horodecki}},
  \bibnamefont{and}
  \bibinfo{author}{\bibfnamefont{R.}~\bibnamefont{Horodecki}},
  \emph{\bibinfo{title}{Mixed-state entanglement and distillation: is there a
  {`}bound{'} entanglement in nature?}}, \bibinfo{journal}{Physical Review
  Letters} \textbf{\bibinfo{volume}{80}}, \bibinfo{pages}{5239}
  (\bibinfo{year}{1998}), \eprint{quant-ph/9801069}.

\bibitem[{\citenamefont{Rudolph}(2002)}]{Rudolph02}
\bibinfo{author}{\bibfnamefont{O.}~\bibnamefont{Rudolph}},
  \emph{\bibinfo{title}{Further results on the cross norm criterion for
  separability}} (\bibinfo{year}{2002}), \eprint{quant-ph/0202121}.

\bibitem[{\citenamefont{Chen and Wu}(2003)}]{CW02}
\bibinfo{author}{\bibfnamefont{K.}~\bibnamefont{Chen}} \bibnamefont{and}
  \bibinfo{author}{\bibfnamefont{L.~A.} \bibnamefont{Wu}},
  \emph{\bibinfo{title}{A matrix realignment method for recognizing
  entanglement}}, \bibinfo{journal}{Quantum Information and Computation}
  \textbf{\bibinfo{volume}{3}}, \bibinfo{pages}{193} (\bibinfo{year}{2003}),
  \eprint{quant-ph/0205017}.

\bibitem[{\citenamefont{Rudolph}(2003)}]{Rudolph02c}
\bibinfo{author}{\bibfnamefont{O.}~\bibnamefont{Rudolph}},
  \emph{\bibinfo{title}{Some properties of the computable cross norm criterion
  for separability}}, \bibinfo{journal}{Physical Review A}
  \textbf{\bibinfo{volume}{67}}, \bibinfo{pages}{032312}
  (\bibinfo{year}{2003}), \eprint{quant-ph/0212047}.

\bibitem[{\citenamefont{Horodecki et~al.}(2002)\citenamefont{Horodecki,
  Horodecki, and Horodecki}}]{HHH02}
\bibinfo{author}{\bibfnamefont{M.}~\bibnamefont{Horodecki}},
  \bibinfo{author}{\bibfnamefont{P.}~\bibnamefont{Horodecki}},
  \bibnamefont{and}
  \bibinfo{author}{\bibfnamefont{R.}~\bibnamefont{Horodecki}},
  \emph{\bibinfo{title}{Separability of mixed quantum states: linear
  contractions approach}} (\bibinfo{year}{2002}), \eprint{quant-ph/0206008}.

\bibitem[{\citenamefont{Vidal and Werner}(2001)}]{VW01b}
\bibinfo{author}{\bibfnamefont{G.}~\bibnamefont{Vidal}} \bibnamefont{and}
  \bibinfo{author}{\bibfnamefont{R.~F.} \bibnamefont{Werner}},
  \emph{\bibinfo{title}{A computable measure of entanglement}}
  (\bibinfo{year}{2001}), \eprint{quant-ph/0102117}.

\bibitem[{\citenamefont{Fan}(2002)}]{Fan02}
\bibinfo{author}{\bibfnamefont{H.}~\bibnamefont{Fan}}, \emph{\bibinfo{title}{A
  note on separability criteria for multipartite state}}
  (\bibinfo{year}{2002}), \eprint{quant-ph/0210168}.

\bibitem[{\citenamefont{Wocjan and Horodecki}(2005)}]{WH05b}
\bibinfo{author}{\bibfnamefont{P.}~\bibnamefont{Wocjan}} \bibnamefont{and}
  \bibinfo{author}{\bibfnamefont{M.}~\bibnamefont{Horodecki}},
  \emph{\bibinfo{title}{Characterization of combinatorically independent
  permutation separability criteria}} (\bibinfo{year}{2005}),
  \eprint{quant-ph/0503129}.

\bibitem[{\citenamefont{Horodecki et~al.}(1996)\citenamefont{Horodecki,
  Horodecki, and Horodecki}}]{HHH96}
\bibinfo{author}{\bibfnamefont{M.}~\bibnamefont{Horodecki}},
  \bibinfo{author}{\bibfnamefont{P.}~\bibnamefont{Horodecki}},
  \bibnamefont{and}
  \bibinfo{author}{\bibfnamefont{R.}~\bibnamefont{Horodecki}},
  \emph{\bibinfo{title}{Separability of mixed states: necessary and sufficient
  conditions}}, \bibinfo{journal}{Physics Letters A}
  \textbf{\bibinfo{volume}{223}}, \bibinfo{pages}{1} (\bibinfo{year}{1996}),
  \eprint{quant-ph/9605038}.

\bibitem[{\citenamefont{Bru{\ss} and Peres}(2000)}]{BP99}
\bibinfo{author}{\bibfnamefont{D.}~\bibnamefont{Bru{\ss}}} \bibnamefont{and}
  \bibinfo{author}{\bibfnamefont{A.}~\bibnamefont{Peres}},
  \emph{\bibinfo{title}{Construction of quantum states with bound
  entanglement}}, \bibinfo{journal}{Physical Review A}
  \textbf{\bibinfo{volume}{61}}, \bibinfo{pages}{30301} (\bibinfo{year}{2000}),
  \eprint{quant-ph/9911056}.

\bibitem[{\citenamefont{D{\"u}r and Cirac}(2000)}]{DC99}
\bibinfo{author}{\bibfnamefont{W.}~\bibnamefont{D{\"u}r}} \bibnamefont{and}
  \bibinfo{author}{\bibfnamefont{J.~I.} \bibnamefont{Cirac}},
  \emph{\bibinfo{title}{Classification of multi{--}qubit mixed states:
  separability and distillability properties}}, \bibinfo{journal}{Physical
  Review A} \textbf{\bibinfo{volume}{61}}, \bibinfo{pages}{042314}
  (\bibinfo{year}{2000}), \eprint{quant-ph/9911044}.

\bibitem[{\citenamefont{Bennett et~al.}(1999)\citenamefont{Bennett,
  Di{V}incenzo, Mor, Shor, Smolin, and Terhal}}]{BDMSST99}
\bibinfo{author}{\bibfnamefont{C.~H.} \bibnamefont{Bennett}},
  \bibinfo{author}{\bibfnamefont{D.~P.} \bibnamefont{Di{V}incenzo}},
  \bibinfo{author}{\bibfnamefont{T.}~\bibnamefont{Mor}},
  \bibinfo{author}{\bibfnamefont{P.~W.} \bibnamefont{Shor}},
  \bibinfo{author}{\bibfnamefont{J.~A.} \bibnamefont{Smolin}},
  \bibnamefont{and} \bibinfo{author}{\bibfnamefont{B.~M.}
  \bibnamefont{Terhal}}, \emph{\bibinfo{title}{Unextendible product bases and
  bound entanglement}}, \bibinfo{journal}{Physical Review Letters}
  \textbf{\bibinfo{volume}{82}}, \bibinfo{pages}{5385} (\bibinfo{year}{1999}),
  \eprint{quant-ph/9808030}.

\bibitem[{\citenamefont{{\.{Z}}yczkowski
  et~al.}(1998)\citenamefont{{\.{Z}}yczkowski, Horodecki, Sanpera, and
  Lewenstein}}]{ZHSL98}
\bibinfo{author}{\bibfnamefont{K.}~\bibnamefont{{\.{Z}}yczkowski}},
  \bibinfo{author}{\bibfnamefont{P.}~\bibnamefont{Horodecki}},
  \bibinfo{author}{\bibfnamefont{A.}~\bibnamefont{Sanpera}}, \bibnamefont{and}
  \bibinfo{author}{\bibfnamefont{M.}~\bibnamefont{Lewenstein}},
  \emph{\bibinfo{title}{On the volume of the set of mixed entangled states.}},
  \bibinfo{journal}{Physical Review A} \textbf{\bibinfo{volume}{58}},
  \bibinfo{pages}{883} (\bibinfo{year}{1998}), \eprint{quant-ph/9804024}.

\bibitem[{\citenamefont{Sloane}()}]{sloan}
\bibinfo{author}{\bibfnamefont{N.~J.~A.} \bibnamefont{Sloane}},
  \emph{\bibinfo{title}{The on-line encyclopedia of integer published
  electronically at:{\newline}
  {\href{http://www.research.att.com/~njas/sequences/index.html}{http://www.re%
search.att.com/\symbol{126}njas/sequences/}.}}}

\end{thebibliography}

\end{document}